\documentstyle{article}
\input psfig
\def\vev#1{{\left\langle #1 \right\rangle}}
\def\d{{\rm d}}
\def\bfr{{\bf r}}
\def\bfx{{\bf x}}
\def\bfk{{\bf k}}

\def\bfU{{\bf U}}
\def\pmb#1{\setbox0=\hbox{#1}%
\kern-.025em\copy0\kern-\wd0
\kern.05em\copy0\kern-\wd0
\kern-.025em\raise.0433em\box0}
\def\iras{{\it IRAS\/}}
\def\etal{{\it et al.\/}}
\def\kms{\ifmmode {\rm \ km \ s^{-1}}\else $\rm\, km \ s^{-1}$\fi}
\def\mpc{\ifmmode {\,h^{-1}\rm Mpc}\else$\,{h^{-1}\rm Mpc}$\fi}
\def\kpc{\ifmmode {\,h^{-1}\rm kpc}\else$\,{h^{-1}\rm kpc}$\fi}

\def\ltsima{$\; \buildrel < \over \sim \;$}
\def\simlt{\lower.5ex\hbox{\ltsima}}
\def\gtsima{$\; \buildrel > \over \sim \;$}
\def\simgt{\lower.5ex\hbox{\gtsima}}
\def\bfv{{\bf v}}
\def\grad{{\bf \nabla }}
\begin{document}

\title{The Large-Scale Velocity Field}

\author{Michael A. Strauss\\Princeton University Observatory}

\maketitle

\section{Bulk Flows as a Cosmological Probe}
\label{sec:importance}

Hubble's Law, now spectacularly confirmed by the work of
\cite{Lauer92}, \cite{Perlmutter96}, and \cite{Riess96}, tells us that the distances of galaxies are
proportional to their observed recession velocities, at least at low
redshifts: 
\begin{equation} 
cz = H_0 r\quad.
\label{eq:Hubble} 
\end{equation}
However, this is not exactly correct.  Galaxies have {\it peculiar
velocities\/} above and beyond the Hubble flow indicated by
Eq.~(\ref{eq:Hubble}).  We denote the peculiar velocity $\bfv(\bfr)$ at every
point in space; the observed redshift in the rest frame of the Local
Group is then:
\begin{equation} 
cz = H_0 r + \hat\bfr\cdot\left(\bfv(\bfr) - \bfv({\bf 0})\right)\quad,
\label{eq:cz-r}
\end{equation}
where the peculiar velocity of the Local Group itself is $\bfv({\bf
0})$, and $\hat \bfr$ is the unit vector to the galaxy in question.
In practice, we will measure distances in units of \kms, which means
that $H_0 \equiv 1$, and the uncertainties in the value of $H_0$
discussed by Freedman and Tammann in this volume are not an
issue. Thus measurements of redshifts $cz$, and of
redshift-independent distances via standard candles, yield
estimates of the radial component of the velocity field.

What does the resulting velocity field tell us?  On scales large
enough that the rms density fluctuations are small, the equations of
gravitational instability can be linearized, yielding a direct
proportionality between the divergence of the velocity field and the
density field at late times \cite{Peebles80}, \cite{Peebles93}:
\begin{equation} 
\grad \cdot \bfv(\bfr) = -\Omega^{0.6} \delta(\bfr)\quad .
\label{eq:linear-theory} 
\end{equation}

This equation is easily translated to Fourier space: 
\begin{equation} 
i \bfk \cdot \tilde \bfv(\bfk) = -\Omega^{0.6} \tilde \delta(\bfk)\quad,
\label{eq:linear-Fourier} 
\end{equation}
which means that if we define a {\it velocity\/} power spectrum
$P_v(k) \sim \vev{\tilde\bfv^2(\bfk)}$ in analogy with the usual density
power spectrum $P(k)$, we find that 
\begin{equation} 
P_v(k) = \Omega^{1.2}k^{-2} P(k)\quad .
\label{eq:power-compare} 
\end{equation}

There are several immediate conclusions that we can draw from
this. Peculiar velocities are tightly coupled to the {\it matter} density
field $\delta(\bfr)$.  Therefore, peculiar velocities are a probe of
the {\it matter} power spectrum; any bias of the distribution of
galaxies relative to that of matter is not an issue.  Moreover,
Eq.~(\ref{eq:power-compare}) shows that it is in principle easier to
probe large spatial scales with peculiar velocities than with the
density field, because of the two extra powers of $k$ weighting for the
velocity power spectrum.  

 Eq.~(\ref{eq:linear-theory}) shows that a comparison of the velocity
field with the {\it galaxy\/} density field $\delta_{\rm gal}$ allows a test
of gravitational instability theory.  However, in order to do this,
one must assume a relation between the galaxy density field (which is
observed via redshift surveys) and the mass density field (which does
the gravitating).  The simplest and most common assumption (other than
simply assuming the two are identical) is that they are proportional
({\it linear biasing}), i.e., $\delta_{\rm gal} = b\, \delta$. If this
is the case, then we can rewrite Eq.~(\ref{eq:linear-theory}) to give:
\begin{equation} 
\grad \cdot \bfv(\bfr) = -{{\Omega^{0.6}}\over b} \delta_{\rm gal}(\bfr) 
\equiv -\beta\, \delta_{\rm gal}(\bfr)\quad.
\label{eq:v-delta-gal} 
\end{equation}
Thus if the observed density and velocity field are consistent with
one another and Eq.~(\ref{eq:v-delta-gal}), we can hope to measure
$\beta$.  This, and other approaches to $\Omega$ via peculiar
velocities are reviewed in Dekel's contribution to this volume;
cf., the reviews by \cite{Dekel94} and \cite{SW}. 

\section{The Predicted Large-Scale Velocity Field: The Theorist's
View}
\label{sec:theory-view}
\begin{figure}
\centerline{\psfig{file=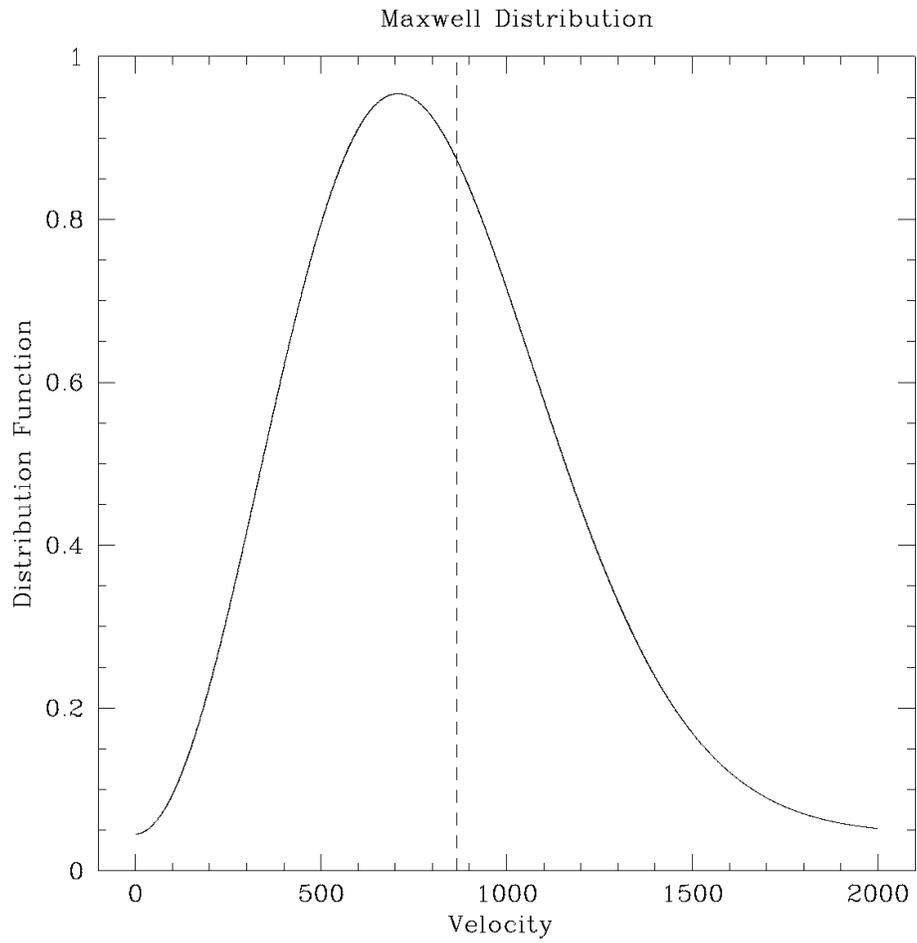,width=12cm}}
\caption{The Maxwellian distribution function of expected bulk flows,
on a scale on which the rms value is 866 \kms. Notice how broad the
distribution is.}
\label{fig:maxwell}
\end{figure}

If a theorist is asked what the large-scale velocity field should look
like, she will use the results derived above to calculate the expected
amplitude of the bulk flow $v(R)$ averaged over a scale $R$:
\begin{equation} 
\vev{v(R)^2} = {{\Omega^{1.2}}\over {2\,\pi^2}} \int\! \d k\, P(k)
\widetilde W^2(kR)\quad,
\label{eq:v-P(k)} 
\end{equation}
where $\widetilde W$ is the Fourier Transform of the smoothing window.
It is straightforward to calculate this quantity as a function of
scale for any given power spectrum (cf., Fig.~9 of \cite{Strauss95}),
but going the other way is more difficult.  If the phases of the
Fourier modes of the density field are random, then each component of
the velocity field has a Gaussian distribution, which means that
$v(R)$ has a {\it Maxwellian} distribution; Fig.~\ref{fig:maxwell}
reminds us just how broad such a distribution is.  Therefore, a single
measurement of the bulk flow on large scales gives us a relatively weak
handle on the power spectrum.

  How then can we constrain the observed power spectrum with
observations of the velocity field?   Under the random phase
hypothesis, the
velocity field is given by a multi-variate Gaussian, whose covariance
matrix can be calculated directly from the power spectrum
(\cite{Gorski88}; \cite{Gorski89}; \cite{Feldman94};
\cite{Watkins95}; \cite{Jaffe95}; \cite{Zaroubi96}).  The velocity
correlation function is then a tensor with elements given by:
\begin{equation} 
\pmb{$\Psi$}_{\mu\nu}(\bfr) \equiv \vev{v_\mu(\bfx)v_\nu(\bfx + \bfr)} =
\Psi_\perp(r)\delta_{\mu\nu} + \left[\Psi_\Vert(r) -
\Psi_\perp(r)\right]\hat r_\mu \hat r_\nu\quad,
\label{eq:vel-correlation} 
\end{equation}
where, in linear perturbation theory,
\begin{equation} 
\Psi_{\perp,\Vert}(r) = {{\Omega^{1.2}}\over {2\,\pi^2}} \int\! \d k\, P(k)
K_{\perp,\Vert}(kR)\quad,
\label{eq:two-psis} 
\end{equation}
and $K_{\perp,\Vert}(x)$ are appropriate combinations of spherical
Bessel functions. 

Thus, {\it if measurements of peculiar velocity for different galaxies
are independent}, then the covariance matrix between radial peculiar
velocities $u_i,u_j$ of two galaxies $i$ and $j$ separated by a distance $r$
is given by:
\begin{equation} 
C_{ij} = \hat\bfr_i^\dagger\pmb{$\Psi$}(r)\hat\bfr_j + (\Delta u)^2
\delta_{ij}\quad,
\label{eq:vel-covariance} 
\end{equation}
where the second term on the right-hand side contains the effects of measurement errors.
This allows one to write down a simple expression for the likelihood of
observing peculiar velocities of a given set of $N$ galaxies, given a
power spectrum:
\begin{equation} 
{\cal L} = [(2\,\pi)^N \det(C)]^{-1/2} \exp\left(-{1 \over 2}
\sum_{i,j}^N u_i C_{ij}^{-1}u_j\right)\quad.
\label{eq:likelihood} 
\end{equation}
This has been applied most recently by \cite{Zaroubi96}, who used the
Mark III peculiar velocity compilation of
\cite{Willick95},\cite{Willick96a},\cite{Willick96b} to constrain the
power spectrum (see \cite{Kolatt95} for an independent determination
of the power spectrum from the same data using the statistics of the
smoothed $\grad \cdot \bfv$).  If they do not apply the constraint of
the COBE \cite{Bennett96} normalization, they find the best-fit CDM
models to have a $\Gamma \equiv \Omega h = 0.5 \pm 0.15$, which
interestingly calls for {\it less} large-scale power than has been
implied, e.g., by large-scale redshift surveys. 

  It is not clear, however, whether the error contributions to the
covariance matrix (Eq.~\ref{eq:vel-covariance}) are purely diagonal.
In particular, if there is an error in the assumed distance indicator
relation which is used to measure peculiar velocities, or if the
distance indicator relation is calibrated from the dataset itself as
in \cite{Lauer94}, covariance is introduced between all peculiar
velocities, introducing off-diagonal terms throughout.  The effect of
this on the determination of the power spectrum remains an area for
further work. 

\section{The Predicted Large-Scale Velocity Field: The Observer's 
View}
\label{sec:observe-view}

\begin{figure}
\centerline{\psfig{file=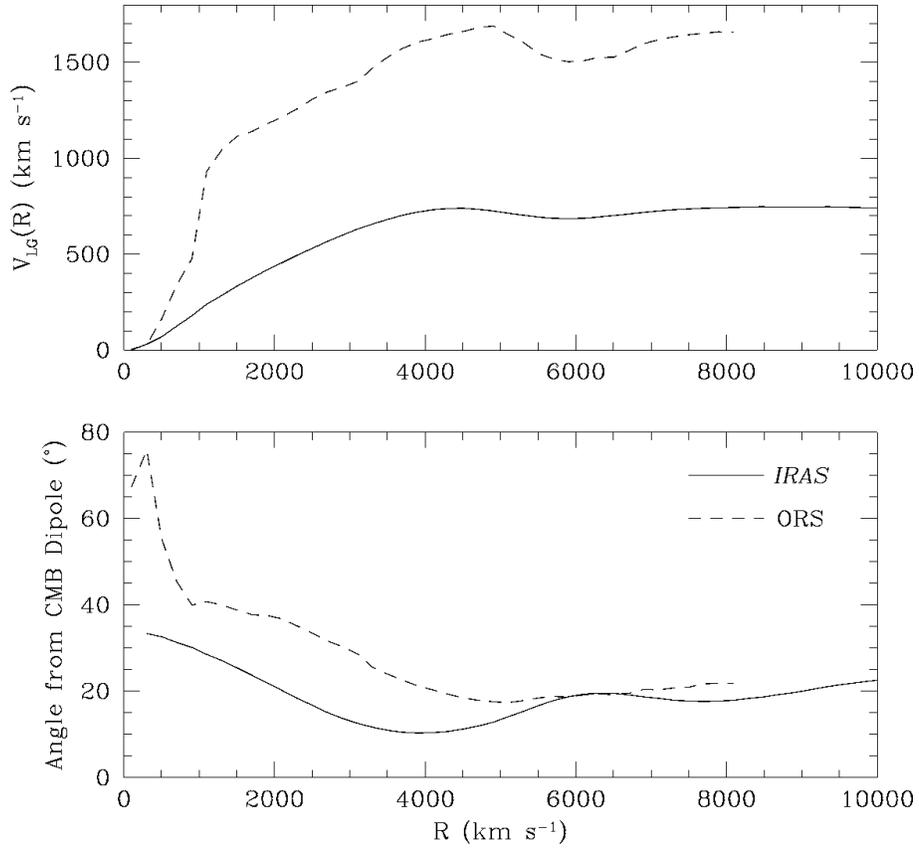,width=12cm}}
\caption{The amplitude (upper panel) and direction relative to the CMB
dipole (lower panel) of the gravitational dipole of two surveys, the
\iras\ 1.2 Jy survey (solid lines) and the Optical Redshift Survey
(dashed lines).  Notice the apparent convergence of the dipole in
both cases beyond roughly 4000 \kms\ (although the two differ quite a
bit in amplitude).  Notice also that the ORS is not quite as deep
as the \iras\ sample, and therefore the dipole calculation is cut off sooner.
}
\label{fig:dipole}
\end{figure}

The observed distribution of galaxies from redshift surveys gives a
prediction for the large-scale components of the bulk flow via the
integral version of Eq.~(\ref{eq:v-delta-gal}) . In
particular, we observe from observations of the dipole anisotropy of
the CMB (e.g., \cite{Kogut93}) that the Local Group is moving with
a velocity of $627 \pm 22 \kms$ towards $l = 276^\circ,\ b =
+30^\circ$ (with $3^\circ$ errors in each angular coordinate); this is
indeed by far the most accurately measured peculiar velocity we know.
One can {\it predict\/} this peculiar velocity from the observed
galaxy distribution to be:
\begin{equation} 
\bfv_{LG} = {\beta \over 4\,\pi n_1} \sum_{{\rm galaxies}\ i}
{W(r_i) \, \hat
\bfr_i \over \phi(r_i)\, r_i^2}\quad,
\label{eq:dipole} 
\end{equation}
where $\phi(r)$ is a selection function, to correct for the decrease
in density of galaxies as a function of distance in a flux-limited
sample, and $W(r)$ is a window function with cutoffs at large and
small scales (cf., \cite{Strauss92}).  The cutoff is needed at large
distance because any flux-limited sample has only finite depth, and
therefore the dipole one calculates is missing contributions from
large scales (\cite{Juszkiewicz90}; \cite{Lahav90}; \cite{Peacock92}).
Indeed, one might think that the difference between the observed and
predicted motion of the Local Group would be a direct measure of large-scale
components of the velocity field.  Fig.~\ref{fig:dipole} shows the
growth of the amplitude and direction of the predicted motion
$\bfv_{LG}(R)$ as a function of the redshift $R$ out to which galaxies
are included in the sum, for two redshift surveys: the \iras\ 1.2 Jy
redshift survey (\cite{Fisher95}; cf., \cite{Strauss92}), and the
Optical Redshift Survey (\cite{Santiago95}; \cite{Santiago96}).
Interestingly, the two curves have a very different amplitude, which
has interesting things to tell us about the relative bias of \iras\
and optically-selected galaxies, but discussing that would get us too
far afield.  For the moment, notice that both curves seem to converge
to a constant value (both amplitude and direction) for $cz > 4000
\kms$, implying that there is little contribution on larger scales.
This in turn would imply that the sphere of radius 4000 \kms\ is at
rest.

  Unfortunately, things are not so simple.  First, as Juszkiewicz
\etal\ \cite{Juszkiewicz90} pointed out, the difference between the
true peculiar velocity and $\bfv_{LG}(R)$ depends on the position of
the center of mass of the sample out to $R$:
\begin{equation} 
\bfv_{LG}(R = \infty) = \bfv_{LG}(R) + \bfv_{bulk}(R) - {1 \over 3}
\beta\, r_{\rm center\ of\ mass}\quad,
\label{eq:dipole-growth} 
\end{equation}
where $\bfv_{bulk}(R)$ is the quantity we're interested in in the
current context, the mean bulk flow of the sphere out to radius
$R$.  One can calculate the rms position of the center of mass of a
sample given a power spectrum from linear theory; one finds another
integral over the power spectrum like Eq.~(\ref{eq:v-P(k)}), although
with a different smoothing kernel.  This term is quite small for small
values of $R$, but becomes comparable to the expected rms bulk flow
for values of $R$ above $5000\kms$ or so \cite{Strauss95}, and
indeed, for the \iras\ 1.2 Jy sample, $r_{\rm center\ of\ mass}$ is of the
order of 250 \kms\ for an outer radius of 10,000 \kms. 

More important than this, however, are all the additional effects
which cause the quantity in Eq.~(\ref{eq:dipole}) to differ from the
theoretical ideal.  Non-linear effects, shot noise, assuming the
incorrect value of $\beta$ (which of course we don't know) and the
smoothing on small scales all will contribute to the difference
between the observed and predicted motion of the Local Group
\cite{Strauss92}.  The most pernicious effect, however, was pointed
out by 
\cite{Kaiser87}.  With a redshift survey, one is measuring the density
field in redshift space.  However, as Eq.~(\ref{eq:cz-r}) makes clear,
this differs from the bulk flow in real space by the effects of
peculiar velocities, and to the extent that the peculiar velocity
field shows coherence (which of course is what we're trying to get a
handle on here), Eq.~(\ref{eq:dipole-growth}) is systematically
biased.  In particular, if one's estimate of the velocity of the Local
Group itself is off (e.g., if one doesn't correct for the $\bfv({\bf
0})$ term in Eq.~(\ref{eq:cz-r}) at all), the positions of {\it all\/}
galaxies in the sample are affected in a dipolar way, clearly
affecting the predicted motion of the Local Group, and the apparent
convergence, or lack thereof, of $\bfv_{LG}(R)$.  Strauss \etal\ \cite{Strauss92}
find that with their best correction of the density field for peculiar
velocities, the \iras\ dipole indeed seems to converge quite nicely,
but even then, there is a very intriguing, large contribution to the
dipole (albeit at the 2$\sigma$ level) between 17,000 and 20,000
\kms. It will be very interesting to see whether this contribution
remains with the just completed PSCZ survey of \iras\ galaxies to 0.6
Jy (cf., Efstathiou, this volume). 

\section{The Measurement of Bulk Flows}
\label{sec:bulk-measurement}
The quantity we hoped to get a handle on from the convergence of the
density dipole, Eq.~(\ref{eq:dipole}), is the average peculiar
velocity of galaxies within a sphere of radius $R$ centered on the
Local Group.  One approach is to measure it directly from a full-sky
peculiar velocity survey.  It is one of the lowest-order statistics
one might imagine measuring from a peculiar velocity sample, but it is
maximally sensitive to systematic errors in observations between
different areas of the sky. 

 In particular, most peculiar velocity surveys carried out to date
have been done over a relatively limited area of the sky. If they are
calibrated externally (as they usually are), zero-point differences
between the calibrators and the sample will give rise to false bulk flow
measurements. Moreover, Malmquist bias can give an artificial signature
of outflow \cite{Bothun90}.

  To avoid these problems, we would like to have measurements of
peculiar velocities over the full sky.  We have already made reference
above to the Mark III dataset of \cite{Willick95}, \cite{Willick96a},
\cite{Willick96b}, which combines one $D_n-\sigma$ \cite{Faber89}, and
six Tully-Fisher \cite{Mould91}; \cite{Mould93}; \cite{Courteau92},
\cite{Aaronson82} (cf., \cite{Tormen95}); \cite{Willick91};
\cite{Mathewson92} peculiar velocity samples.  A great deal of work
has been done to make these datasets consistent by matching them where
they overlap.  The bulk flow of the resulting full-sky sample has been
calculated by \cite{Courteau93}, and more recently by Dekel \etal\ (in
preparation).  I describe the latter calculation here.

\begin{figure}
\centerline{\psfig{file=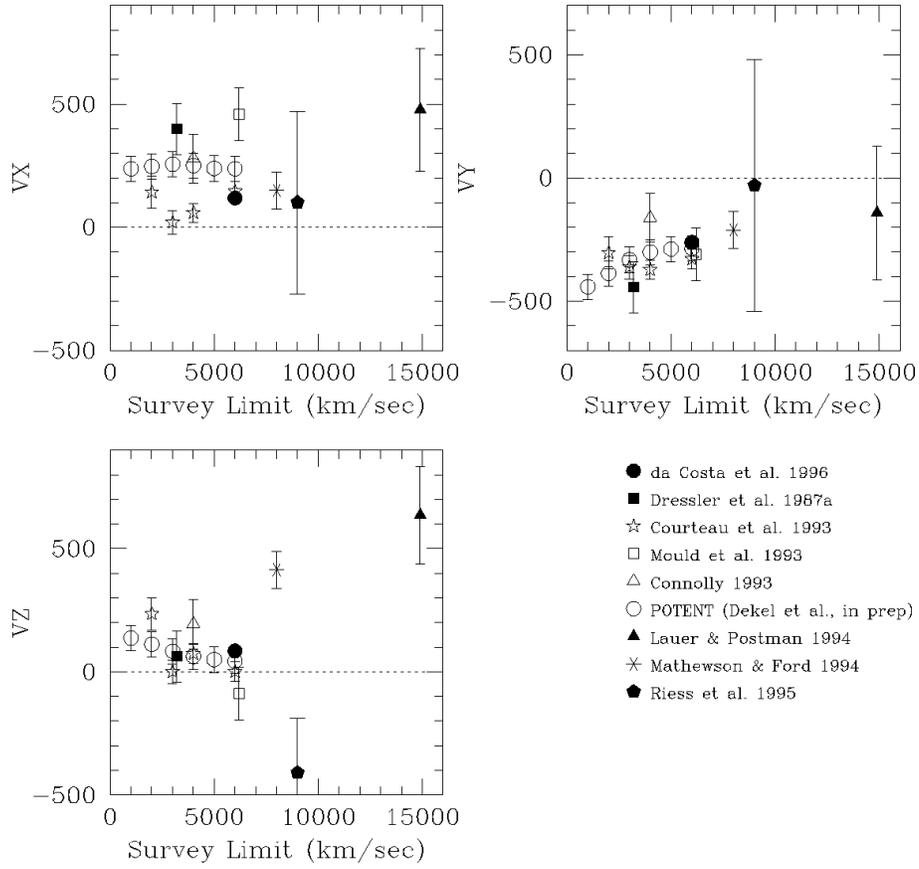,width=12cm}}
\caption{Determinations of
the bulk flow of galaxies on various scales from the literature. The
three panels give the components of the quoted bulk flows along the
Galactic $X$, $Y$, and $Z$ directions in \kms, as a function of the
depth of the various surveys. Error bars are as quoted by each paper,
and do not take into account the covariance between the different
directions (i.e., due to error ellipsoids whose principal axes are not
aligned with the Galactic Cartesian directions). Adapted from
Postman 1995.} 
\label{fig:postman}
\end{figure}

  The peculiar velocity data are noisy and sample the field sparsely
and inhomogeneously.  The data can be smoothed if one assumes that the
velocity field is derivable from a potential; this allows the
calculation of a unique three-dimensional velocity field from
observations of radial peculiar velocities (the POTENT method;
\cite{Dekel90}; \cite{Dekel94}; \cite{daCosta96}; Dekel, this volume).
Calculating the bulk flow is then straightforward, and the results are
shown in Fig.~\ref{fig:postman}.

This approach has the advantage that the bulk flow that is calculated
is close to the theorist's ideal, the volume-weighted bulk
flow. Indeed, the straight fit of individual peculiar velocities in a
sample to a bulk flow will not be equivalent to the volume-weighted
bulk flow, both because of clustering within the sample (cf., the
discussion in \cite{Strauss95}) and because of the increasing peculiar
velocity errors with distance (cf., \cite{Kaiser88}).

  However, measuring a bulk flow on large scales requires tremendous
control over systematic photometric errors.  Indeed, a 0.10 mag
difference in the photometric zero-points of the Mark III sample from
one end of the sky to another would translate into an artificial 300
\kms\ bulk flow at 6000 \kms\ from the Local Group.  Davis \etal\
\cite{Davis96} have carried out a multipole comparison of the Mark III
peculiar velocity field with that predicted from the \iras\ 1.2 Jy
redshift survey, and found that there are indeed discrepancies between
the two fields beyond 4000 \kms\ of roughly 300 \kms\ amplitude.  It
remains unclear whether this is the signature of the gravitational
influence of dark matter whose distribution has nothing to do with
that of galaxies, a 
sign that peculiar velocities are not wholly due to the process of
gravitational instability, or more prosaically, that there are
systematic errors in the Mark III data which are unaccounted for.

In the latter regard, Fig.~\ref{fig:postman} compares various
determinations of the bulk flows of galaxies within spheres centered
on us that have been published in the literature.  This figure is an
updated version of one shown by \cite{Postman95a}.  Error
bars are as given by each author, and do not take into account any
misalignment between the error ellipsoids and the the Cartesian axes
chosen.  The current confused situation is reflected in the large number of
non-overlapping error bars in this figure. However, note that the bulk
flow within 6000 \kms\ of Dekel \etal\ (from the POTENT analysis of
the Mark III data) and of \cite{daCosta96} (from the POTENT-like
analysis of the Giovanelli \etal\ data; cf., Giovanelli in this
volume) are in excellent agreement, despite almost completely
independent data (they do share the Mathewson \etal\
\cite{Mathewson92} data in common).  It will be very interesting to
see if they agree this well shell-by-shell.

\section{Full-Sky Peculiar Velocity Surveys}
\label{sec:full-sky}
Given the uncertainty introduced by possible zero-point differences
between the samples making up the Mark III dataset, how can the bulk
flows on large scales best be measured? The ideal way is with a
peculiar survey of galaxies covering the entire sky, observed in as
uniform a way as possible.  In particular, the survey should
\begin{itemize}
\item have full-sky, uniform sampling in angle and redshift;

\item have well-defined, simple, and easily modeled selection criteria;

\item use a distance indicator with small intrinsic dispersion;

\item use uniform observing techniques between North and South,
Spring and Fall, with much repeat observations.
\end{itemize}

There are a number of surveys just completed or in progress now which
approach this ideal. In particular:

\begin{itemize}
\item Roth \cite{Roth94} and Schlegel \cite{Schlegel95} have carried
out a Tully-Fisher study of a full-sky volume-limited sample of 140
\iras\ galaxies to 4000 \kms. A bulk flow analysis has not yet been
done, although the data have been compared to the \iras\ predicted
velocity field, and have found consistency for the relatively small
value of $\beta = 0.4$ \cite{Schlegel95}.

\item The EFAR collaboration \cite{Wegner96} has carried out a
$D_N-\sigma$ study of over 700 elliptical galaxies in 84 clusters in
the Hercules-Corona Borealis, and Perseus-Pisces-Cetus directions in
the redshift range $6000 < cz < 15000 \kms$, with the aim of
constraining the velocity fields in these superclusters.

\item Hudson \etal\ (in preparation) are extending the EFAR survey
with measurements of $D_n-\sigma$ distances to 6-10 ellipticals in
those clusters in the Lauer-Postman (1994) sample \cite{Lauer94} with
redshifts $cz < 12,000 \kms$.

\item Fruchter, Moore \& Steidel (in preparation) are doing wide-field
photometry of a subsample of the Lauer-Postman clusters; a fit of the
photometry to a Schechter function then yields a distance. 

\item J. Willick has measured Tully-Fisher distances to 20 spirals in
each of 15 clusters of galaxies over the sky, at $cz \approx 10,000
\kms$.  Analysis is in progress. 

\item Giovanelli \etal\ are measuring Tully-Fisher distances to spiral
galaxies both in the field and in clusters over a large fraction of
the sky to redshifts of 6000 \kms\ and greater; see Giovanelli in this
volume. 
\end{itemize}

There are two further surveys in which I am involved, which I
describe in the following two sections. 

\section{The Bulk Flow of Brightest Cluster Galaxies}
\label{sec:warpfire}
Lauer and Postman \cite{Lauer94}, \cite{Postman95b} presented
distances of the Brightest Cluster Galaxies (BCG's) of a sample of 119
Abell \cite{Abell58}, \cite{Abell89} clusters to $cz < 15,000 \kms$.
Following work of \cite{Gunn75} and \cite{Hoessel80}, they found that
the luminosity $L$ of these galaxies within an aperture of radius
10\kpc\ correlated with the logarithmic slope of the surface
brightness profile $\alpha$.  This yields a distance indicator with an
error of $15-20\%$, depending on the value of $\alpha$.  Their sample
was full-sky (or as much so as the zone of avoidance would allow) and
volume-limited, and great effort was taken to obtain and reduce the
data as uniformly as possible.

To their great surprise, the sample showed a strong signature of bulk
flow, with an amplitude of $764 \pm 160 \kms$ \cite{Colless95}, towards
$l = 341^\circ,\ b = +49^\circ$.  This was much larger than one might
expect, given the effective depth of the sample of $\approx 8000
\kms$; indeed, \cite{Feldman94} and \cite{Strauss95}
both showed that a bulk flow with the statistical significance of that
of Lauer-Postman ruled out a whole series of cosmological models at
the $>95\%$ confidence level. 

\begin{figure}
\centerline{\psfig{file=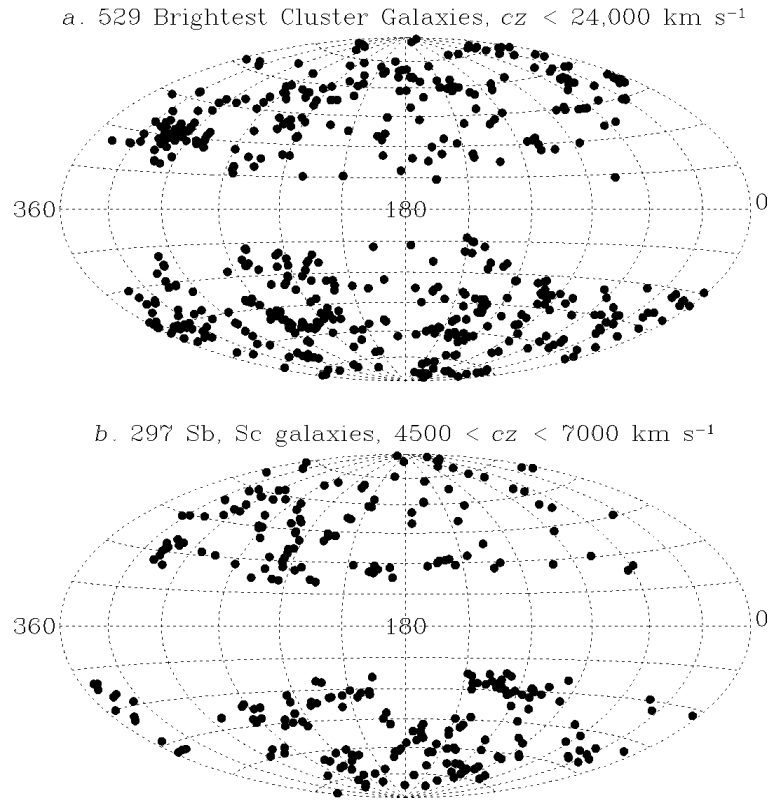,width=10cm}}
\caption{{\it a}. The BCG sample
with $z < 0.08$.  The
substantial region devoid of clusters in the general direction of the
Galactic center is due to the difficulty in finding clusters in
regions of high stellar density, and is a general feature of the Abell
catalogue.
{\it b}. The sky
distribution in Galactic coordinates of galaxies in the Sb, Sc shell
sample at $cz \approx 6000 \kms$.}
\label{fig:skydist}
\end{figure}

As a follow-up to this survey, Tod Lauer, Marc Postman and I are
extending the sample to $cz = 24,000 \kms$.  The sample now consists
of 529 BCG's, an increase of more than a factor of 4 from the original
l19 (the Abell cluster catalog has the beautiful feature of being
volume-limited, at least to moderate redshifts, and this increase in
number of clusters is almost exactly the increase in volume).  The
photometry for this sample is all in hand, and redshifts for all BCG's
are nearly complete.  Barring unseen systematic effects (which we've
worked very hard to minimize), we should be able to measure the bulk
flow on these scales to 130\kms\ or so.  We have also measured
velocity dispersions for the BCG's, with preliminary indications that
this reduces the scatter in the $L-\alpha$ relation, in analogy to the
$D_n-\sigma$ relation. The sky distribution of this sample is shown in
Fig.~\ref{fig:skydist}a.

As Fig.~\ref{fig:postman}, and the controversy that the Lauer-Postman
result have engendered, make clear, the comparison of various
measurements of bulk flows with one another is
non-trivial. 

%To start with, we are not just looking for ``consistency with
%convergence,'' in the sense that we discussed in
%\S~\ref{sec:observe-view}.  That is, we would like to do more than to
%confirm that the peculiar velocity field as measured from some
%specific sample is consistent with zero bulk flow in the CMB frame
%with very large error bars.  Such a statement does not necessarily
%rule out another determination of the bulk flow on similar scales. 

The velocity field has components on all scales; it is not
purely dipolar in nature. The geometry of any given sample couples to
various multipoles of the velocity field (the sparser the sampling is,
the larger the extent to which this is true), and therefore not all bulk flow
measurements measure the same quantity \cite{Watkins95}.
Thus \cite{Riess95} published a bulk flow analysis of 13 Type 1a
supernovae, which appear to be standard candles to an accuracy of
$\sim 5$\% \cite{Riess96}.  Their results were inconsistent with
that found by Lauer \& Postman at the 99\% confidence
level, {\it assuming that the velocity field was describable by a pure
bulk
flow plus small-scale incoherent noise}.  However, the two surveys
sample space really very differently, and therefore are very
differently sensitive to components of the velocity field on scales
smaller than the dipole.  Watkins \& Feldman \cite{Watkins95} calculated the
expectation value of the dot product of the bulk flows each measured,
normalized by the expectation value of each bulk flow separately: 
\begin{equation} 
{\cal C} \equiv {\vev{\bfU^{LP}\cdot\bfU^{RPK}}
\over\left(\vev{\bfU^{LP}\cdot\bfU^{LP}}\vev{\bfU^{RPK}\cdot\bfU^{RPK}}\right)^{1/2}}\quad.
\label{eq:covariance} 
\end{equation}
This quantity, a sort of dimensionless covariance between the two
bulk flow measurements, would be close to unity if these two surveys
were indeed measuring the same quantity.  The results depend on the
power spectrum assumed.  If one assumes ``realistic'' power spectra,
the quantity $\cal C$ is of the order of 10\%, but as mentioned above,
the Lauer-Postman result is inconsistent with most ordinary power
spectra.  Watkins \& Feldman thus also consider a power spectrum with
a huge bump at large scales; in such a model, the relative importance
of small-scale components of the velocity field is reduced, but the
quantity $\cal C$ is still only 35\%. 

\section{Resolving the Discrepancies}
\label{sec:resolution}

Wandering through the halls of astronomy departments around the
country (or even reading preprints on the astro-ph archive), one hears
a lot of interesting statements about the large-scale bulk flow of
galaxies within 6000 \kms:

``The Lauer-Postman result cannot be right; it does not agree with
observed bulk flow measurements at 6000 \kms.''

``The Lauer-Postman result cannot be right; it does not agree with
the fact that the \iras\ dipole appears to have converged by 6000 \kms.''

``The observed bulk flow at 6000 \kms\ from Mark III is inconsistent with
the predictions of the \iras\ redshift survey.''

``The Mark III and da Costa \etal\ \cite{daCosta96} dipoles are inconsistent
with one another at 6000 \kms.''

Clearly, much of the current controversy centers around the bulk flow
at 6000 \kms. St\'ephane Courteau, Marc Postman, Dave Schlegel,
Jeff Willick, and I have started a full-sky Tully-Fisher survey of
galaxies specifically designed to nail down the bulk flow within a
shell centered at 6000 \kms. We have selected 297 Sb-Sc galaxies with
$4500 < cz < 7000 \kms$ with appropriate inclinations and without
morphological peculiarities, from the magnitude-limited full-sky
redshift survey sample of \cite{Santiago96} (we decided against
using \iras\ selection, given the large Tully-Fisher scatter observed
for \iras\ galaxies in \cite{Schlegel95}).  The sky distribution of
this sample is shown in Fig.~\ref{fig:skydist}b.  For each galaxy, we measure
the rotation curve using a long slit for the H$\alpha$ line, and are
doing photometry in the $V$ and $I$ bands.  We have been granted
observing time at Kitt Peak and Cerro Tololo for this survey, and we
hope to finish in one year.  Our estimate is that we will be able to
measure the bulk flow of this shell with an error of 70\kms, with an
error ellipsoid that will be close to isotropic.  We believe that this
survey should resolve much of the controversy that is currently
swirling around this very hot topic. 

\medskip
I would like to acknowledge my collaborators in the various projects I
discuss here: the \iras\ 1.2 Jy and ORS redshift surveys (Marc Davis,
Alan Dressler, Karl Fisher, John Huchra, Ofer Lahav, Bas\'\i lio
Santiago, and Amos Yahil), the POTENT/Mark III analysis (David
Burstein, St\'ephane Courteau, Avishai Dekel, Sandy Faber, and Jeff
Willick), and the two bulk flow projects described above (St\'ephane
Courteau, Tod Lauer, Marc Postman, David Schlegel, and Jeff Willick).
I acknowledge the support of a Fellowship from the Alfred P. Sloan
Foundation.

\end{document}